\documentclass[12pt]{iopart}

\usepackage[T1]{fontenc}
\usepackage[utf8]{inputenc}
\usepackage{graphicx}
\usepackage{array}
\usepackage{natbib}
\usepackage{sidecap}
\usepackage{caption}
\usepackage{subfig}
\usepackage{dcolumn}
\usepackage{bm}
\usepackage{float}
\usepackage{lineno}
\usepackage{wrapfig}
\usepackage{rotating}
\usepackage{booktabs}
\usepackage{geometry}
\usepackage[symbol]{footmisc}

\begin{document}

\title{A theoretical framework to predict the most likely ion path in particle imaging}
\author{Charles-Antoine Collins-Fekete$^{1,2,3}$, Lennart Volz$^{4,5}$, Stephen K. N. Portillo$^{6}$, Luc Beaulieu$^{1,2}$, Joao Seco$^{3,4,5}$}
\address{$^1$Département de physique, de génie physique et d’optique et Centre de recherche sur le cancer, Université Laval, Québec, Canada \\
$^2$Département de radio-oncologie et CRCHU de Québec, CHU de Québec, QC, Canada \\
$^3$Department of Radiation Oncology, Francis H. Burr Proton Therapy Center Massachusetts General Hospital (MGH), Boston, MA, USA \\
$^4$Deutsches Krebsforschungszentrum Heidelberg, Baden-Württemberg, DE \\
$^5$University of Heidelberg, Department of Physics and Astronomy Heidelberg, Baden-Württemberg, DE \\
$^6$Harvard-Smithsonian Center for Astrophysics, Cambridge, MA, USA \\
}
\ead{charles-antoine.collins-fekete.1@ulaval.ca}


\submitto{\PMB}

\begin{abstract}
In this work, a generic rigorous Bayesian formalism is introduced to predict the most likely path of any ion crossing a medium between two detection points. The path is predicted based on a combination of the particle scattering in the material and measurements of its initial and final position, direction and energy. The path estimate's precision is compared to the Monte Carlo simulated path. Every ion from hydrogen to carbon is simulated in two scenarios to estimate the accuracy achievable: one where the range is fixed and one where the initial velocity is fixed. In the scenario where the range is kept constant, the maximal root-mean-square error between the estimated path and the Monte Carlo path drops significantly between the proton path estimate (0.50 mm) and the helium path estimate (0.18 mm), but less so up to the carbon path estimate (0.09 mm). In the scenario where the initial velocity is kept constant, helium have systematically the minimal root-mean-square error throughout the path. As a result, helium is found to be the optimal particle for ion imaging. 
\end{abstract}

%================================================
\section{Introduction}
%================================================
%% introduce charged particles imaging
Proton radiography (pRad) and proton computed tomography (pCT) were first proposed by \cite{cormack_representation_1963} and later experimentally proved (\cite{cormack_quantitative_1976}). However, non-deflected photon tomography soon proved to be much more efficient and straightforward, and research in proton imaging halted. 

\ \\
%%introduce the interest of proton and charged particles imaging for therapy
Recently, this research emerged again with the advent of proton therapy. The proton therapy planning system requires knowledge of the proton stopping power within the patient, which is measured by proton tomography. As of now, this quantity is clinically obtained through a conversion from X-ray tomography Hounsfield units \citep{schneider_calibration_1996}. Such a process introduces large uncertainties in planning and reduces the flexibility and advantages of proton treatment \citep{matsufuji_relationship_1998,schaffner_precision_1998,chvetsov_influence_2010,yang_comprehensive_2012}. It has been proven that single-event pCT could help reduce the uncertainty by directly measuring the proton stopping power in the patient \citep{zygmanski_measurement_2000}. Moreover, proton imaging possesses several clinical and diagnostic qualities. It has a higher density resolution, a significantly lower noise level, and lower dose to the patient (\cite{schulte_density_2005,depauw_sensitivity_2011}) than in the conventional X-ray CT imaging. Finally, pCT suffers from different artifacts than X-ray CT (\cite{depauw_sensitivity_2011}). However, one of the major problem encountered in pCT is the lower spatial resolution compared to X-ray CT. 

\ \\
%%introduce the problem of the most likely path
The multiple deflections a proton suffers throughout its path, known as multiple Coulomb scattering (MCS), greatly reduce the spatial resolution of the images acquired. Consequently, the conventional X-ray tomographic algorithm struggles when using unaltered proton radiographies to reconstruct the pCT and the extracted images are of poor quality. To resolve the problem of MCS, accurate proton path estimate methods have been proposed. Of those, the most likely path (MLP) algorithm is the most widely applied (\cite{schneider_multiple_1994,williams_most_2004,schulte_maximum_2008,erdelyi_comprehensive_2009}). It is a method to calculate the proton path given position and direction information as well as the proton beam scattering formulated from the Fermi-Eyges \citep{rossi_cosmic-ray_1941,eyges_multiple_1948} scattering equation. Starting from a different perspective, \cite{fekete_developing_2015} proposed a phenomenological approach to retrieve the proton path from a fit of the cubic spline direction magnitude that best reproduces the Monte Carlo estimated path. Nevertheless, every approach relies on a sophisticated proton by proton detection system \citep{schulte_conceptual_2004,bashkirov_proton_2007,hurley_water-equivalent_2012} to acquire precise entry and exit position/direction data and energy loss. Still, the MLP algorithm is limited by the inherent uncertainty associated with the MCS and can not resolve the proton path with high precision, leading to tomographic images of inferior spatial resolution than X-ray CT. 

\ \\
On the other hand, heavier ions suffer less from MCS due to smaller average angular deviations and are viable candidates to acquire high-quality tomographic images. However, more sophisticated accelerators are required to produce a beam of heavier particles with a minimal energy to cross a clinically relevant distance \citep{pedroni_accelerators_1993,lomax_charged_2009,kitagawa_review_2010,owen_current_2016}. Moreover, as of now no trajectory estimate has been proposed to extract the ions MLP.

\ \\
%%introduce the innovation in this work and how solve the problem of charged particle imaging
This work first presents a generalized formalism that strictly follows the Bayesian theory to estimate the most likely trajectory of an ion. The proposed formalism is demonstrated to encompass the prior non-rigorous Bayesian formalism by \cite{schulte_maximum_2008}, itself based on work by \cite{schneider_multiple_1994} and \cite{williams_most_2004}, and replicate the phenomenological cubic spline path (CSP) prediction made by \cite{fekete_developing_2015}. Furthermore, the formalism strict Bayesian definition allows for future extensions with an example shown in this work. The new formalism is  used to investigate the accuracy of the path estimate for heavier ions crossing a medium using the accuracy of the proton's path estimate as a comparison. The MLP maximal root mean square (RMS) error to the Monte Carlo path is investigated for every ion up to carbon.

%================================================
\section{Most likely path formalism}
%================================================
%%make a quick summary of what will be presented here
The Bayesian formalism used to express the MLP of a proton particle crossing a medium is first defined. The definitions of the likelihoods are extracted from the Fermi-Eyges theory, leading to a simple and intuitive MLP equation. Then, the formalism is extended to every ion.

%%--------------------------------------------------
\subsection{Bayesian formalism}

\begin{figure}[!ht]
\centering
\includegraphics[width=0.9\textwidth]{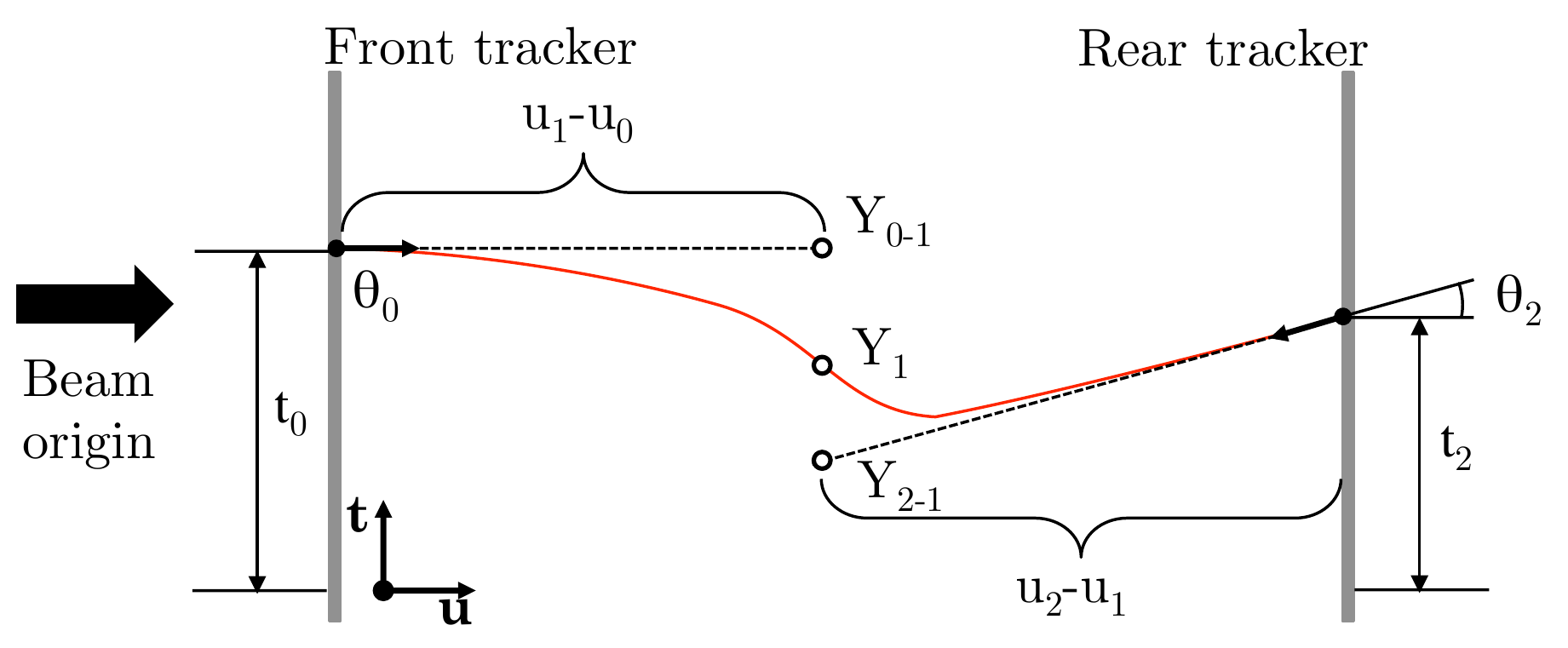}
\caption{Representation of the Bayesian formalism with two likelihoods as suggested in this work. Both measurement points are propagated up to the lateral depth of the reconstructed point. The MLP algorithm represents the average of these points weighted by their inverse scattering, function of the distance from their respective detector.}
\label{fig:schema_MLPpriors}
\end{figure}

%%Schulte formalism
This formalism starts from the same premises that have been laid down by \cite{schulte_maximum_2008} when it was proposed to use Bayesian theory to extract the MLP of a proton crossing a medium between two detection planes. Briefly, in Bayesian probability theory, the posterior probability of an event is proportional to the product of the prior expectation of the probability of the event with the likelihood of current observations. Given a model with unknown parameters that describes a system, the prior represents the expectations for these parameters before making measurements, while the likelihood represents the probability of measurements given some specific system parameters. 

\ \\
In this work, the longitudinal position in the phantom is given by the parameter $u$ ($u_0$ labels the position of the front detector, $u_2$ the position of the rear detector). The lateral position is given by the parameter $t$. The entrance and exit angles are represented by the parameters $\theta_0$ and $\theta_2$ and defined relative to the longitudinal axis. The 2D parameter vector $Y$ represents the measurements made at the entrance and exit plane ($Y_0$ and $Y_2$), or the sought proton parameter at depth $u_1$~($Y_1$). The parameter vector is defined as Equation \ref{eq:systemparameter}. The concepts introduced here are visualized in Figure~\ref{fig:schema_MLPpriors}.

\begin{eqnarray}
 Y = \left( \begin{array}{c} t \\ \theta\end{array}\right)
\label{eq:systemparameter}
\end{eqnarray}

\ \\
To calculate the MLP, \cite{schulte_maximum_2008} first define the prior as the probability of finding a proton at a parameter vector $Y_1$ given an entrance parameter vector $Y_0$, denoted $\mathcal{L}$($Y_1 |$entry data). The likelihood is set as the probability of finding an exit proton with parameter vector $Y_2$ given the parameter vector $Y_1$ ($\mathcal{L}$(exit data$|Y_1$)). By combining both, they then compute the posterior probability of finding a proton at longitudinal position $u_1$ given the exit data (Equation \ref{eq:schulte_likelihood}). The posterior is then maximized by setting the derivative of the logarithm of the posterior to zero. The proportional term in Equation \ref{eq:schulte_likelihood} comes from the fact that the normalization in Bayesian formalism is not defined here. It will however be of no importance when considering the derivative of the logarithm of the posterior. 

\begin{eqnarray}
\mathcal{L}(Y_1|\textrm{exit data})\propto \mathcal{L}(\textrm{exit data}|Y_1)\mathcal{L}(Y_1|\textrm{entry data}) 
\label{eq:schulte_likelihood}
\end{eqnarray}

%% Problem in Schulte formalism based on their wrong assumption of the Bayesian formalism
\ \\
This definition of the prior does not follow rigorously the Bayesian theory and complicates the derivation of the MLP. The prior is calculated from the propagation of the entry parameter vector $Y_0$ to the parameter vector $Y_1$, which means it requires the entry measurement to have already been performed. However, in Bayesian theory, the prior is usually known before the measurements are acquired and should not rely on them. There is no fundamental reason why the entry measurement should be used. This definition set a non-rigorous base that may complicates further development of the MLP theory.

%%introduce our formalism
\ \\ 
In this work, a strict Bayesian formalism is proposed which greatly simplifies the calculations. The new formalism is expressed as the combination of two likelihoods (defined by the notation $\mathcal{L}$) on the measurements $Y_0$ and $Y_2$ with a prior on $Y_1$ (defined by the notation $\pi$) (Equation \ref{eq:charles_likelihood}).

\begin{eqnarray}
\mathcal{P}(Y_1|Y_0,Y_2) \propto \mathcal{L}(Y_0 | Y_1)\mathcal{L}(Y_2 | Y_1)\pi(Y_1)
\label{eq:charles_likelihood}
\end{eqnarray}

\ \\
The variables introduced in the formalism proposed by \cite{schulte_maximum_2008} are mostly used and redefined here. The Fermi MCS theory \citep{rossi_cosmic-ray_1941} with Eyges solution \citep{eyges_multiple_1948} is used to parameterize the probability of finding a particle originating from a narrow beam at a certain depth given a lateral and angular deviation. The formalism is expressed as a bivariate Gaussian function with the covariance matrix defined by the Eyges moments ($A_n$). The Eyges moments for every ions are defined using Highland integral scattering power \citep{kanematsu_alternative_2008}, shown in Equation \ref{eq:highlandionspecies}.
\begin{eqnarray}
A_n(u,z)= \left(1+\frac{1}{9}ln\left(\int_0^u\frac{du'}{X_0(u')} \right) \right)\sqrt[]{\int_0^u \left(\frac{E_0 z}{pv(u')}\right)^2\frac{(u-u')^n du'}{X_0(u')} }
\label{eq:highlandionspecies}
\end{eqnarray}

\ \\
The term $X_0(u)$ represents the radiation length of the material at depth u, and the empirical constant $E_0$ =14.1 [MeV/c] was used as in  \cite{kanematsu_alternative_2008}$^ {\footnotemark}$\footnotetext{The original Highland constant was 17.5 MeV \citep{highland_practical_1975}. However, the 14.1 MeV constant is referred to as the Highland's constant and used as the standard in most literature \citep{kanematsu_alternative_2008}.} to best represent the scattering of ions crossing a medium. The term $pv$ represents the momentum function relative to the depth in the material, specific to the investigated ion. It can be described through the energy function of the particles throughout the medium. In this work, the Monte Carlo simulations allowed us to generate the $pv(u)$ function and calculate the integral directly.

\ \\
The bivariate Gaussian covariance matrix is defined by the first three Eyges moments (Equation \ref{eq:covariancematrix}).

\begin{eqnarray}
\Sigma_0 = \left( \begin{array}{cc} A_2 &  A_1 \\ A_1 & A_0\end{array} \right) = \left( \begin{array}{cc} \sigma_t^2& \sigma_{t\theta}^2\\ \sigma_{t\theta}^2 & \sigma_{\theta}^2\end{array} \right) 
\label{eq:covariancematrix}
\end{eqnarray} 

\ \\
The likelihood can be expressed using the Fermi-Eyges distribution by calculating the probability of finding a particle originating from the measurement point at a depth $u$ given a lateral $t$ and angular deviation $\theta$. To do so, the measurement point must be propagated up to the reconstruction point $u$. The small angle approximation is used within this work for the entry and the exit angle ($\sin\theta \approx \theta$). This leads to the definition of the transvection matrices $R_0$ and $R_2$.

\begin{equation}
R_0 = \left(\begin{array}{cc} 1 & u-u_0 \\ 0 & 1\end{array}\right), \qquad R_2 = \left(\begin{array}{cc} 1 & u-u_2 \\ 0 & 1\end{array}\right)
\label{eq:propagationmatrices}
\end{equation}

\ \\
The system parameter that represents the combined lateral and angular deviation is defined by ($Y_1-R_0Y_0$). Thus, the likelihood of the observation of the entrance parameter vector (and equivalently from the exit parameter vector) is expressed as in Equation \ref{eq:priorentrance}.

\begin{eqnarray}
\mathcal{L}(Y_0|Y_1) = exp\left( -\frac{1}{2}(Y_1-R_0Y_0)^T\Sigma^{-1}_0(Y_1-R_0Y_0) \right)
\label{eq:priorentrance}
\end{eqnarray}

\ \\
The posterior is defined by the product of both likelihoods and the prior. For the initial simulation, a flat prior is used $\pi(Y_1) = 1$. However, prior knowledge (\textit{e.g.} X-ray CT ) could be used to improve the MLP precision. The posterior definition is shown in Equation \ref{eq:totchisquare}.

\begin{eqnarray}
\mathcal{P}(Y_1|Y_0,Y_2) \propto &\exp(-\frac{1}{2}((Y_1 -R_0Y_0)^T\Sigma_0^{-1}(Y_1-R_0Y_0)+\nonumber \\ &(Y_1 -R_2Y_2)^T\Sigma_2^{-1}(Y_1 -R_2Y_2))
\label{eq:totchisquare}
\end{eqnarray}

\ \\
The value that maximizes the posterior is found by taking the first derivative of \textit{log($\mathcal{P}$)} with respect to the parameter vector $Y_1$. The derivative of the log posterior is set to zero, leading to Equation \ref{eq:mlp}.

\begin{eqnarray}
Y_{MLP} = \left(\Sigma_0^{-1} + \Sigma_2^{-1} \right)^{-1}\left(\Sigma_0^{-1}R_0Y_0 +\Sigma_2^{-1}R_2Y_2  \right) 
\label{eq:mlp}
\end{eqnarray}

%% Discuss briefly the meaning of the equation retrieved from our most likely path.
\ \\
Equation \ref{eq:mlp} demonstrates a very simple representation of the $Y_{MLP}$.~The MLP is a linear combination of both measurements propagated to the reconstruction point (expressed by the terms $R_0Y_0$ and $R_2Y_2$) and weighted by the inverse of their scattering covariance matrices. The scattering described within these matrices increases with the distance from the measurements. This sets up a balance in the importance granted to each measurement as a function of the particle energy loss throughout it's path and the radiation length of the material crossed. 

\ \\
The $Y_{MLP}$ equation can be further reduced. For simplicity, let us define two weight matrices (Equation \ref{eq:matricesAB}):

\begin{eqnarray}
(\Sigma_0^{-1} + \Sigma_2^{-1})^{-1}\Sigma_0^{-1} = \left( \begin{array}{cc} A & B \\ C & D\end{array} \right) \nonumber \\ (\Sigma_0^{-1} + \Sigma_2^{-1})^{-1}\Sigma_2^{-1} = \left( \begin{array}{cc} E & F \\ G & H\end{array} \right)
\label{eq:matricesAB}
\end{eqnarray} 

\ \\
The formalism that predicts the lateral deviation can then finally be expressed in a compact form (Equation \ref{eq:mlp_compact}).
\begin{eqnarray}
t_{MLP}= t_0 (A) + \theta_0(A(u_1-u_0) + B) + t_2(E) + \theta_2(E(u_1-u_2) +F)
\label{eq:mlp_compact}
\end{eqnarray}
\ \\
This equation bears a striking resemblance to the natural cubic spline formalism for the most likely path presented in \cite{fekete_developing_2015} (Equation \ref{eq:fekete_2015_mlp}).

\begin{eqnarray}
t(u) =& t_0(2\kappa^3-3\kappa^2+1) +\theta_0(\Lambda_0(u_2-u_0)(\kappa^3-2\kappa^2+\kappa)) +\nonumber \\ &t_2(3\kappa^2-2\kappa^3) + \theta_2(\Lambda_2(u_2-u_0)(\kappa^3-\kappa^2))
\label{eq:fekete_2015_mlp}
\end{eqnarray}
Where $\kappa\in[0,1]$ represents the fractional range defined as $\frac{u-u_0}{u_2-u_0}$.

%%--------------------------------------------------
\subsection{Derivation of the most likely path for every ion}

The scattering power for an ion with charge \textit{z}  is defined following the Highland formulation (Equation \ref{eq:highlandionspecies}) \citep{kanematsu_alternative_2008}. The local ion's momentum can be retrieved from the proton momentum scaled by the ratio of the ion and proton mass (Equation \ref{eq:pv_mass}).
\begin{eqnarray}
pv_p(E) = \frac{E_p(E_p+2m_pc^2)}{(E_p+m_pc^2)} \leftrightarrow E_p = \gamma m_pc^2 \nonumber \\
pv_z(E) = \frac{m_p}{m_z} pv_p(E) = Apv_p(E)
\label{eq:pv_mass}
\end{eqnarray}

\ \\
Where $E_p$ is the proton kinetic energy, $m_pc^2$ is the proton rest energy and $\gamma$ is the Lorentz factor for relativistic particles. Equation \ref{eq:pv_mass} combined with the Highland formulation for ions (Equation \ref{eq:highlandionspecies}) reveals that the Eyges moments for ions scale by the ratio of their mass and charge (Equation \ref{eq:highland_scale}).
\begin{equation}
A_n(u,z)  = \left(\frac{z}{A}\right)^2A_n(u)
\label{eq:highland_scale}
\end{equation}

\ \\
Although this conclusion was reached through the Highland formulation, it holds true for other known scattering formalisms such as the \O verh\aa s-Schneider, the Fermi-Rossi and the differential Highland calculation \citep{gottschalk_scattering_2010}. Based on the definition of the scattering matrices ($\Sigma_0$ and $\Sigma_2$), the posterior $\mathcal{P}$ that defines the ion's path (Equation \ref{eq:totchisquare}) also varies with the ratio of its mass/charge to the proton mass/charge.

\ \\ 
Furthermore, as expressed earlier, the MLP is found by setting the derivative of the log posterior ($\nabla log\mathcal{(P)}$) to zero. Therefore, the mass and charge constants in front of Equation \ref{eq:mlp} cancel out. This leads to a key observation that the maximum-likelihood formalism is the same for any ion it does not depend on the charge of the particles or their mass. 

%%--------------------------------------------------
\subsection{Gaussian variance of the most likely path for every ion}
The MLP is calculated as the mean value of the normal distribution described by the Gaussian scattering of protons coming from two measurements. It is possible to describe the variance function by seeking the standard deviation of the same distribution. This error function is found by taking the inverse of the log-likelihood second derivative.

\begin{eqnarray}
\epsilon_{t_1\theta_1} =  \nabla^2 log(\mathcal{P})^{-1} = 2(\Sigma_0^{-1} + \Sigma_2^{-1})^{-1}
\end{eqnarray}

\ \\
Which returns an error matrix for the variance in the lateral displacement distribution, the angular displacement distribution and the correlated angular-lateral displacement distribution. Although the MLP formalism does not change as a function of the ion used, the Gaussian standard deviation around the MLP will be scaled by the factor $\left(\frac{z}{A}\right)^2$.

%%--------------------------------------------------
\subsection{Energy loss function of different ions and related most likely path precision}

The MLP precision depends both on the initial energy and the energy loss. For the same initial energy per unit mass (and therefore the same velocity) considering low energy loss, the scattering should be equivalent for different ions. However, when the medium length increases, the energy loss of heavier ions is greater and the MLP estimate becomes less precise. The Bethe-Bloch electronic stopping power with density and shell correction terms is represented in Equation \ref{eq:Bichsel}. 
\label{sec:wepr}
\begin{equation}
SP(I,E) =  \rho
\frac{4\pi e^4}{m_ec^2\beta^2} z^2 \frac{Z}{A} 
\left\{ 
 \ln
 \left[ \frac{2m_e c^2 \beta^2}{(1-\beta^2)}\right] -\beta^2  
 -\ln I - \frac{C}{Z}  -\frac{\delta}{2}  
 \right\}
\label{eq:Bichsel} 
\end{equation}
\ \\
Here $\beta$ represents the particles velocity in units of the speed of light $c$, $z$ is the particles atomic number. $Z$, $A$ and $\rho$ are the target materials atomic number, mass number and density respectively. $I$ is the targets ionization potential, $C$ the density and $\delta$ the shell correction term. $m_e$ is the electrons mass and $e$ its charge.

\ \\
For the same velocity, the ratio of the ions to the proton electronic stopping power is characterized by $\frac{SP_z(I,E)}{SP_p(I,E)} = z^2$. Given a lighter ion such as helium, the velocity loss will be smaller and the path estimate will therefore be more precise.

%================================================
\section{Simulations and validations}
%================================================

%%--------------------------------------------------
\subsection{Monte Carlo Simulations}

Simulations were done using the Geant4 Monte Carlo code version 10.2.1 \citep{agostinelli_geant4simulation_2003}. Only primary particles were recorded. The standard processes include energy loss and straggling, multiple Coulomb scattering based on Lewis theory \citep{goudsmit_multiple_1940} using the Urban model \citep{urban_multiple_2006},  parameterized interactions with nuclei and electrons and elastic/inelastic ion interactions from Geant4 ion dedicated packages \citep{lechner_validation_2010}. In precise terms, the following physics lists were enabled: the standard electromagnetic option 3 for higher accuracy of electrons, ions and ion tracking without a magnetic field, the ions elastic model and the binary ion models both for elastic and inelastic collisions. Cuts were set to 0.1 mm.
Every simulation was done using a homogeneous water medium as target which led us to set the radiation length to $X_0$=36.1 cm. 
%%--------------------------------------------------
\subsection{Comparison to the phenomenological cubic spline path}
\begin{figure}[ht] 
\centering
\includegraphics[width=0.85\textwidth]{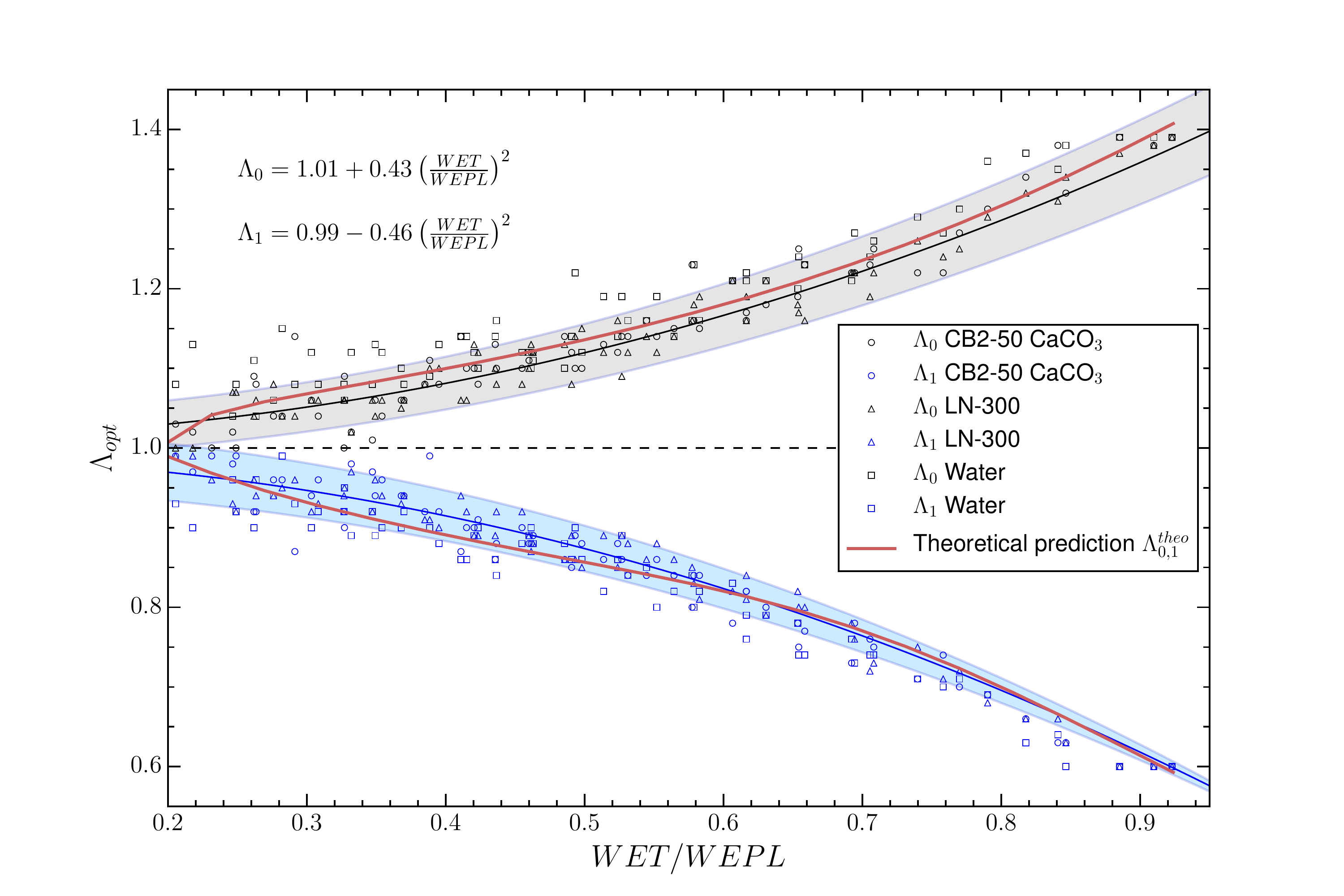}
\caption{Comparison of the theoretical predictions of the factors $\Lambda_0$ and $\Lambda_1$ (red lines) with the $\Lambda_{opt}$ factors that minimize the RMS deviation for different target materials as a function of the WET/WEPL ratio. The black and blue lines represent the parameterization found by \cite{fekete_developing_2015} and the shaded region their uncertainty.}
\label{fig:CSPcomparison}
\end{figure}

Recently, a phenomenological path formalism \citep{fekete_developing_2015} was introduced that makes use of the cubic splines to reproduce the MLP, based on two fit factors $\Lambda_0$ and $\Lambda_1$. As demonstrated previously, the cubic spline equation falls naturally from the proposed Bayesian formalism presented here. Furthermore, it is possible to predict the two fit factors using the $t_{MLP}$ equation derived earlier (Equation \ref{eq:mlp}). 

\ \\
The relation between the phenomenological fit $\Lambda_{0,1}^{opt}$ and the new theoretical predictions is evaluated in Figure \ref{fig:CSPcomparison}. The figure shows the theoretically predicted parameters $\Lambda_{0,1}^{theo}$ as function of the water equivalent thickness/water equivalent path length (WET/WEPL) ratio of the proton energy and thickness crossed. The theoretical curves were obtained employing the Highland scattering \citep{kanematsu_alternative_2008} power to calculate the scattering matrix elements. The theoretical predictions and the phenomenological fit agree within the uncertainty defined by the shaded regions.

%%--------------------------------------------------
\subsection{Estimate of the relative error difference between Monte Carlo and Bayesian path for ions}
The path estimate accuracy for heavier ions up to carbon is investigated for both the CSP and MLP formalism to test the generality of the found relation and validate the conclusions made earlier. In order to do so, the particles (n=$10^6$) were propagated in a water medium. Positions were recorded at 1000 equally spaced points along the phantom for each particle. No detector errors were considered in this work.

\begin{figure}[ht]
\centering
\includegraphics[width=0.75\textwidth]{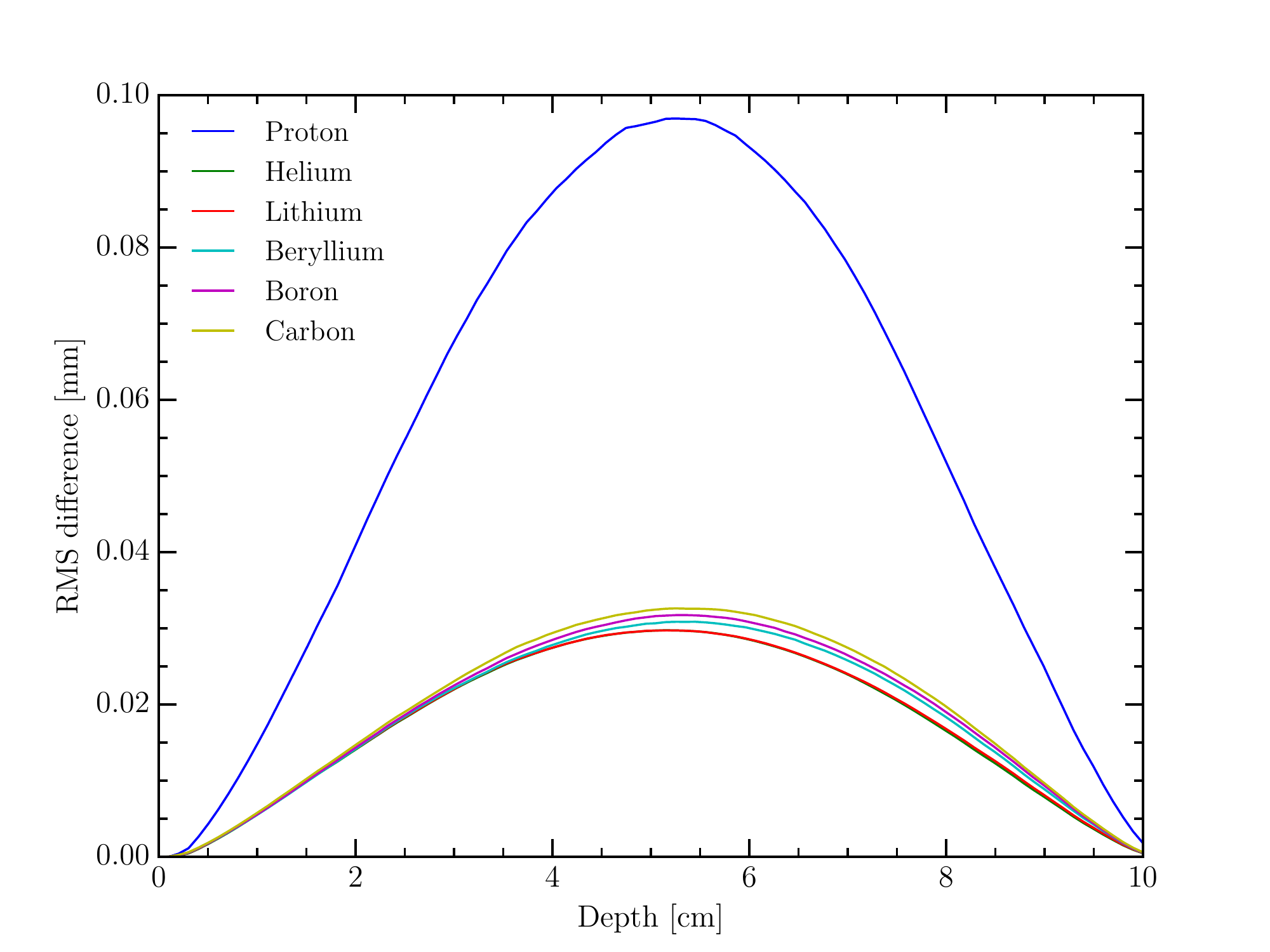}
\caption{RMS deviation between the MLP and the MC paths for each ion up to carbon. In each simulation $10^6$ particles with initial energy of 330 MeV/u were propagated through 10 cm of water.}
\label{fig:RMSComparison}
\end{figure}

%%--------------------------------------------------
\subsubsection{Ion path resolution using a constant initial velocity}

\ \\
The MLP estimates were obtained via the algorithm introduced in the last section. RMS deviations (between CSP and MC as well as MLP and MC) were computed for each ion. Figure \ref{fig:RMSComparison} shows the MLP RMS deviations for each ion up to carbon.

\begin{figure}[!htbp]
  \centering
  \subfloat[330 MeV/u crossing 10 cm of water]{\includegraphics[width=0.50\textwidth]{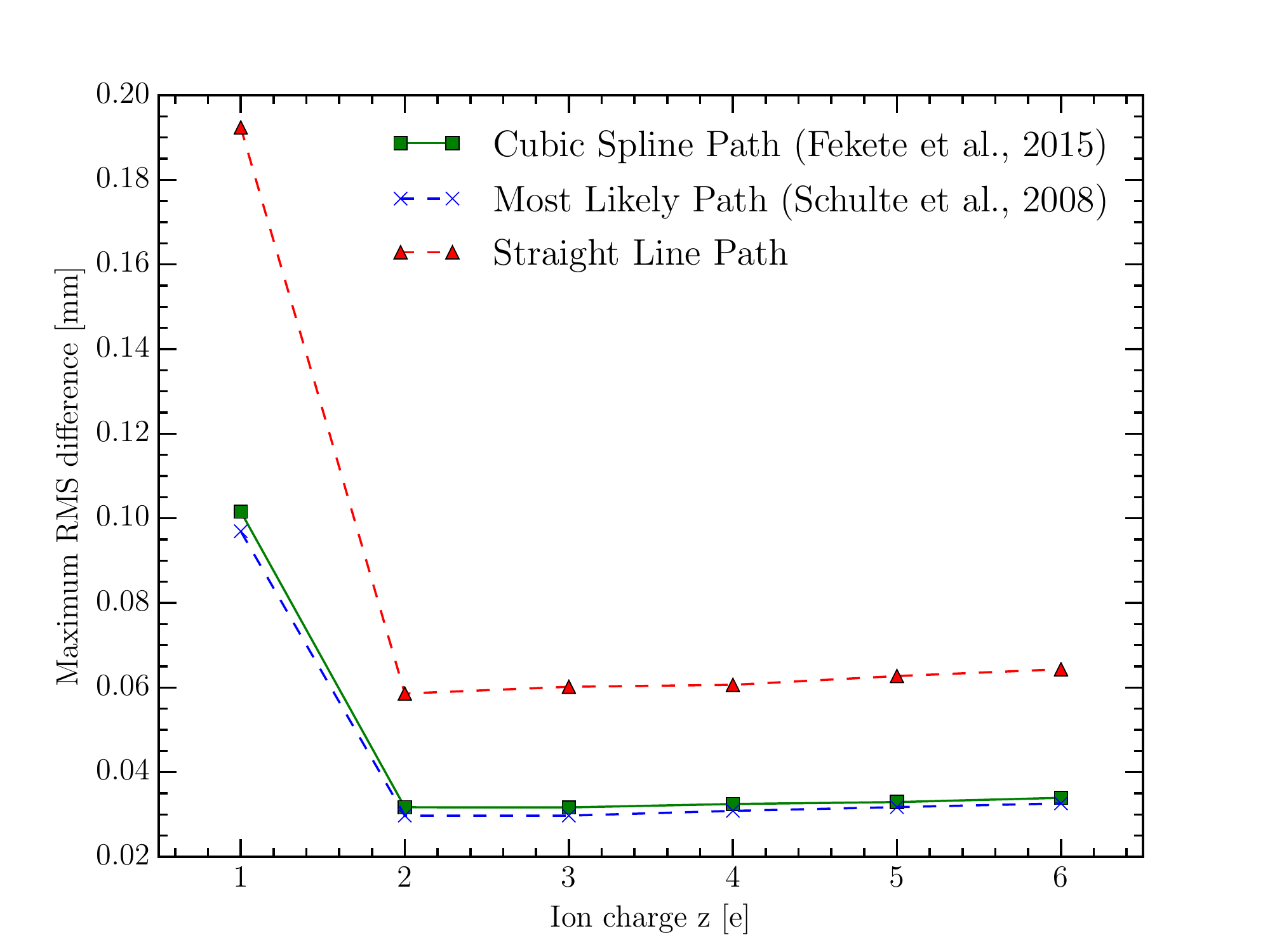}\label{fig:330MeV10CM}}
  \hfill
  \subfloat[330 MeV/u crossing 20 cm of water]{\includegraphics[width=0.50\textwidth]{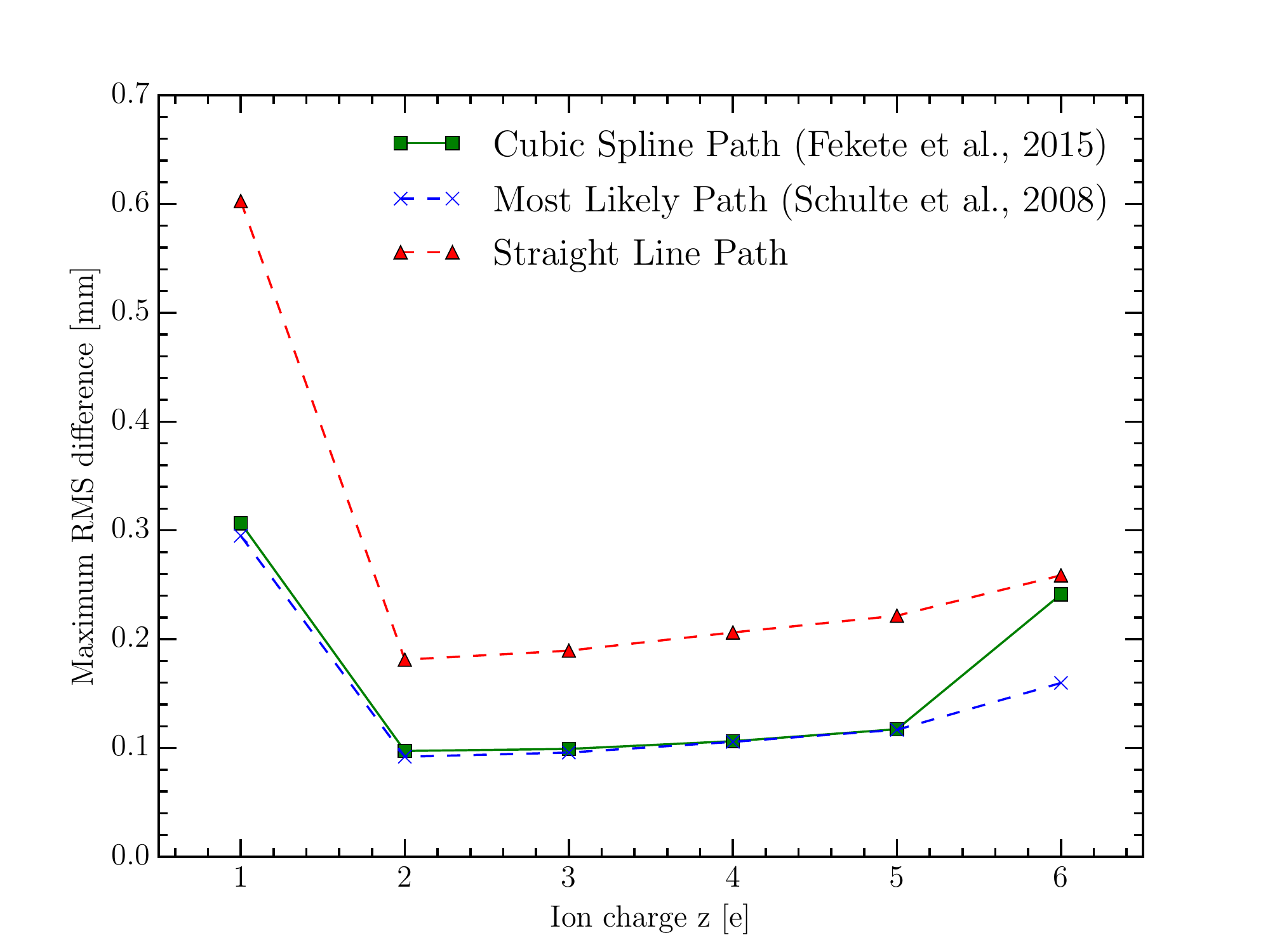}\label{fig:330MeV20CM}}
  \caption{Maximum RMS difference between the estimated path and  the Monte Carlo path for various charged particles and fixed initial velocity. The particles cross 10cm of water (a) and 20 cm of water (b) to represent a situation of low and regular energy loss.}
\end{figure}

\ \\
Figure \ref{fig:330MeV10CM} and \ref{fig:330MeV20CM} shows the maximal RMS deviation between the MLP/CSP path estimate and the Monte Carlo path when the velocity is kept constant for each ions. The path deviation is highest for protons due to the smaller mass to charge ratio, which leads to higher MCS. 

\ \\
The helium has the lowest RMS maximal deviation between the estimated path and the MC path for all scenario with constant entrance energy/mass (Figure \ref{fig:RMSComparison}, \ref{fig:330MeV10CM} and \ref{fig:330MeV20CM}). Finally the path deviation is higher for heavier particles such as carbon relative to helium (Figure \ref{fig:330MeV20CM}) since their velocity loss is greater. 

\begin{figure}[ht]
\centering\includegraphics[width=0.6\textwidth]{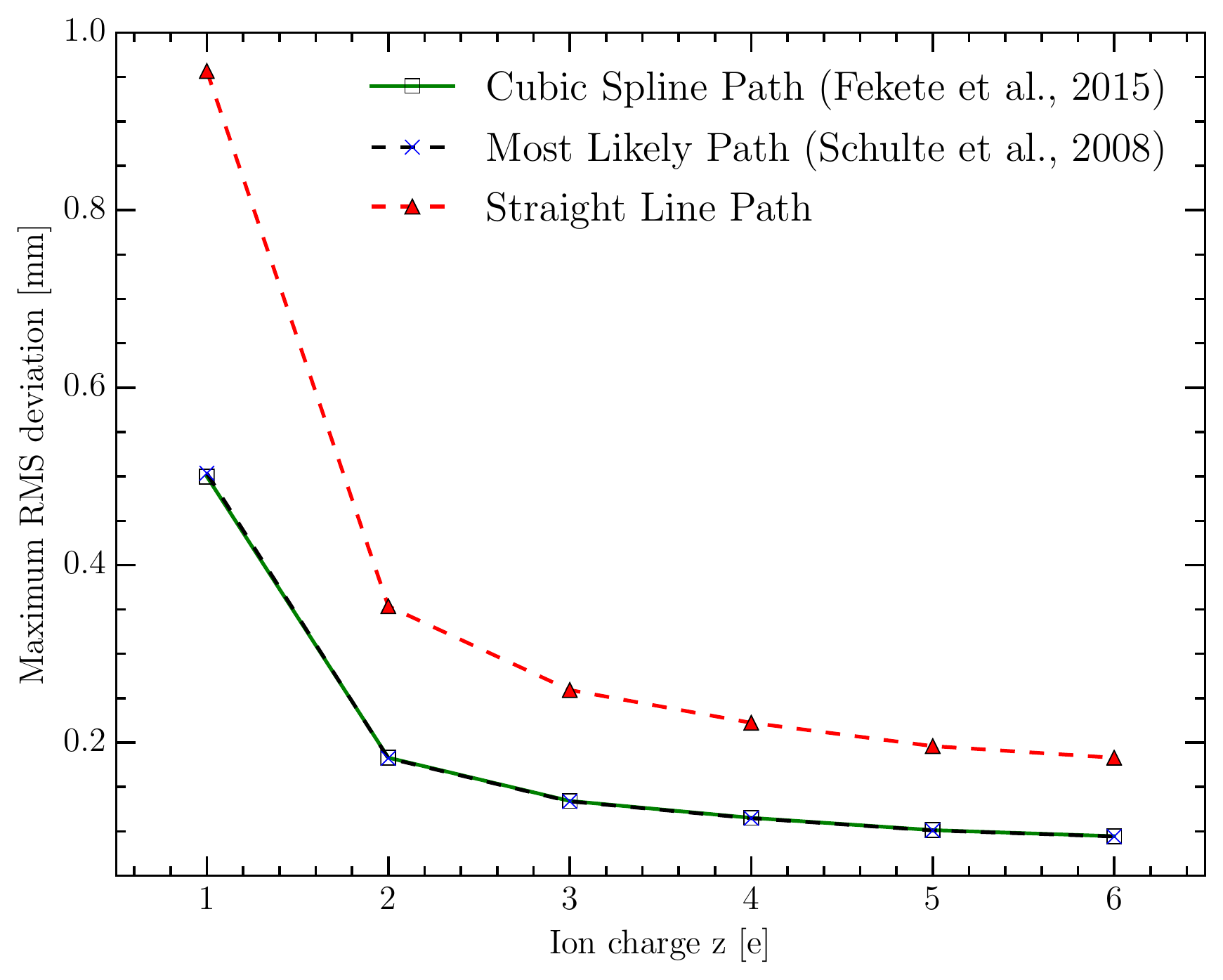}
\caption{Maximum RMS deviation between the CSP path estimate (full green line), the MLP path estimate (black dashed line) and MC as a function of the ion charge z. As a reference, the RMS deviation for a straight line path (red dashed line) is given.}
\label{fig:MaxRMS}
\end{figure}

%%--------------------------------------------------
\subsubsection{Ion path resolution using a constant energy-loss/range}

\ \\
\ \\
Every ions path was simulated through 20 cm of water while ensuring that the range was kept constant. The ion's initial energy was scaled to reproduce a water equivalent range of 26 cm. The required energies and corresponding ranges are given in Table~\ref{tab:EnergyTable}.

\ \\
\begin{table}[h]
\centering
\begin{tabular}{l|c c}
\hline \hline 
ion	&	Energy $\left[MeV/u\right]$	& Range $\left[cm\right]$ \\
\hline 
Proton	&	$200$	&	$25.97$\\
\hline
Helium	&	$200$	&	$26.10$\\
\hline
Lithium	&	$231$	&	$26.03$\\
\hline
Beryllium	& $280$	&	$26.00$\\
\hline
Boron	&	$325$	&	$26.00$\\
\hline
Carbon	&	$386$	&	$26.03$\\
\hline \hline 
\end{tabular}
\caption{Energy per nucleon for different ions corresponding to a range of 26 cm in water. The values were computed using the ICRU report 49 and 73 \citep{deasy_icru_1994,bimbot_stopping_2005,sigmund_errata_2009}}
\label{tab:EnergyTable}
\end{table}
\ \\
For all investigated ions, the MLP and CSP path estimate  are identical and therefore yield the same RMS deviation, as shown in Figure \ref{fig:MaxRMS}. The maximum RMS deviation is slightly shifted towards the rear end of the phantom.~Furthermore, the RMS decreases with increasing ion charge/mass (Figure \ref{fig:MaxRMS}). This decrease is maximal between protons and helium and plateaus for higher z. Additionally, Figure \ref{fig:MaxRMS} illustrates the maximum RMS deviation for a straight line path estimate. The straight-line path also shows a strong decrease from protons to helium, and the maximum RMS values approach those given by the MLP/CSP formalism for higher~$z$. Nonetheless, the maximum RMS deviation for the straight-line path estimate of carbon's trajectory is still approximately twice as high (0.18 vs 0.09 mm), as what is achieved through the MLP/CSP formalism. 

%================================================
\section{Discussion}
%================================================
%First discuss CSP formalism
A  streamlined Bayesian formalism was proposed in this work. It has been shown that the MLP lateral deviation can be expressed in a simple compact equation (Equation \ref{eq:mlp}). Since both the proposed formalism of the MLP and \cite{schulte_maximum_2008} formalism originates from the same scattering theory, the estimated trajectories are identical. However, the approach used here follows strict Bayesian theory which enables a more simple, physically grounded understanding of the MLP: an average path of the projection of the entry and exit position, weighted by the scattering to the point of interest. 

\ \\ 
The phenomenological fit proposed by \cite{fekete_developing_2015} relies on the use of two parameterization factors, $\Lambda_0^{opt}$ and $\Lambda_1^{opt}$. The MLP equation (Equation~\ref{eq:mlp}) yields these factors naturally. It has been shown that both the theoretical prediction and the phenomenological fit agree well within the standard deviations (Figure \ref{fig:CSPcomparison}). In addition, the CSP estimate and the MLP give the same path estimates and RMS deviation for protons with initial energy of 200 MeV (Figure~\ref{fig:RMSComparison}). These arguments make us conclude that the optimized CSP estimate is a simplified and efficient parameterization of the MLP without loss of generality. Furthermore, the CSP algorithm's advantage comes from the fact that it is solely dependent on the ratio of WET and WEPL, rather than on the energy loss at each point in the phantom. 

\ \\
The same MLP equation (Equation~\ref{eq:mlp}) predicts the path for every ion. This formalism has been used to in Figure~\ref{fig:RMSComparison}, verifying the independence of the formalism on ions charge and mass. This conclusion holds true for the CSP formalism as well, as the same RMS deviations between the estimated trajectories and the MC path are computed for all investigated ions.

\ \\
Two scenarios have been investigated to investigate which ion provides the best path estimate. In the first scenario, the same initial velocity is used for all particles (Figure \ref{fig:330MeV10CM} and \ref{fig:330MeV20CM}). In that case, helium is found as the particle with the most precise path estimate regardless of the range crossed.

\ \\
In the second scenario, the energy-loss/range is kept constant (Figure \ref{fig:MaxRMS}) between ions. The gain in accuracy is maximal between proton and helium and becomes less important with increasing ion charge/mass. This is due to the fact that to provide a constant range, the heavier ions particle initial velocity must be drastically increased which creates an unfair comparison. In addition, the use of heavier ions in imaging becomes more complex in practice with increasing mass. The results presented in this work indicate that the best ion for imaging is helium.

\ \\
Throughout this work, a continuous medium of water has been used to evaluate the MLPs accuracy for the ions. However, in a realistic clinical scenario, the medium will be heterogeneous and the MLP should take this factor into account. One way to do so would be to use an heterogeneous function that represents $X_0(u)$ and  $pv(u)$ along the path,  gathered from a prior image ($e.g.$ X-ray CT). Such functions would help the MLP estimate model the path in an heterogeneous phantom. 

\ \\ 
Future work will include simulating the achievable spatial resolution when employing different ions in computed tomography. Furthermore, the effect of nuclear reactions and the resulting secondary particles on the reconstructed images will be investigated as a function of the ion used. Secondary particles are expected to cause artifacts which will reduce the quality of the image. This is expected to increase with the ion charges, with worse artifacts in carbon imaging than helium imaging.

\subsection{Extending the Bayesian framework to account for detectors uncertainty}
The strict Bayesian formalism used here enables the addition of a likelihood that accounts for potential uncertainties related to the detection system. The error induced by the detector can be predicted using a predefined likelihood function of the detection (denoted $\mathcal{L}(\hat{Y_{0}} | Y_0)$) where $\hat{Y_{0}}$ is the position measured by the detection system and $Y_0$ the actual position. In this scenario, the new posterior would be defined as in Equation \ref{eq:likelihood_with_detector_errors}.
\begin{eqnarray}
\mathcal{P}(Y_1|\hat{Y_0},\hat{Y_2}) \propto \mathcal{L}(\hat{Y_0} | Y_0)\mathcal{L}(\hat{Y_2} | Y_2)\mathcal{L}(Y_0 | Y_1)\mathcal{L}(Y_2 | Y_1)\pi(Y_1)
\label{eq:likelihood_with_detector_errors}
\end{eqnarray}
\ \\
This can be used predict the impact of different detectors on the MLP.

%================================================
\section{Conclusion}
%================================================
Based on the Bayesian theory, a mathematical framework for the proton MLP has been presented that leads to a simple representation of the MLP. The MLP was found to be a weighted average of both the entrance and exit measurement points, with the weight being the inverse of the scattering matrices at the reconstruction point. The framework has been proven to be usable for every ion in the same way, emphasizing its generality. The maximum RMS error between the MC simulated path and the estimated path was investigated as a function of the ion used. It was found that the precision in the path estimate is maximal for helium. Furthermore, this ion  has the lowest complexity of production. It is therefore considered the optimal particle for ion imaging.

\bibliographystyle{jphysicsB}
\bibliography{Zotero} % References file
\end{document}